\documentclass{iau}
\usepackage{graphicx}
\usepackage{color}

\title[Parameter Space of Compact Binary Populations] 
{Exploring the Parameter Space of Compact Binary Population Synthesis}

\author[Jim W. Barrett \etal]   
{Jim W. Barrett$^1$, Ilya Mandel$^1$, Coenraad J. Neijssel$^1$, Simon Stevenson$^1$, Alejandro Vigna-G\'omez$^1$}

\affiliation{$^1$School of Physics and Astronomy, University of Birmingham, \\ Birmingham,
B15 2TT, United Kingdom \\
 email: {\tt jbarrett@star.sr.bham.ac.uk} \\[\affilskip]}

\pubyear{2016}
\volume{325}  
\jname{Astroinformatics}
\editors{Massimo Brescia, eds.}
\begin{document}

\maketitle

\begin{abstract}
As we enter the era of gravitational wave astronomy, we are beginning to collect observations
which will enable us to explore aspects of astrophysics of massive stellar binaries which were previously beyond reach.
In this paper we describe COMPAS (Compact Object Mergers: Population Astrophysics and Statistics), a new platform to allow us to deepen our understanding
of isolated binary evolution and the formation of gravitational-wave sources. We describe the
computational challenges associated with their exploration, and present preliminary results on overcoming
them using Gaussian process regression as a simulation emulation technique.
\end{abstract}

\firstsection 
\section{Introduction}

The first detection of gravitational waves
(\cite[Abbott \etal, 2016a]{Abbott_etal2016a})
paves the way for gravitational wave astronomy to become a rich and varied field
of study over the coming decades. During its first observing run, the Advanced Laser Interferometer
Gravitational-Wave Observatory (aLIGO) made two confident detections of the distortions
of space-time caused by the coalescence of a pair of heavy black holes.
(\cite[Abbott \etal, 2016b]{Abbott_etal2016b}) aLIGO is expected to observe many
more black hole mergers, as well as other sources, such as binary neutron star mergers
and black hole-neutron star mergers, over the next few years
(\cite[Abadie \etal, 2010]{Abadie_etal2010};
\cite[Abbott \etal, 2016c]{Abbott_etal2016c}).

The question of how binaries of compact objects form is still an active area of research. Several formation channels
have been proposed, including
isolated binary evolution with a common-envelope phase,
dynamical formation in dense stellar environments or chemically homogeneous evolution.
This paper focuses and the isolated binary evolution channel (see
\cite[Abbott \etal\ 2016b]{Abbott_etal2016b} and references therein for a review).

Compact Object Mergers: Population Astrophysics and Statistics (COMPAS) is a platform currently
under development at the University of Birmingham for the exploration and study of
populations of compact binaries formed through isolated binary evolution
(\cite[Stevenson \etal, 2017]{Stevenson_etal_prep}).
COMPAS is a rapid population synthesis code, designed to be flexible, fast and modular, together with a
collection of tools for a sophisticated statistical treatment of the synthesised populations.

The purpose of binary population synthesis is to use physically intuitive but simplified approximations to detailed
stellar and binary evolution models, so that binaries can be simulated for their entire lifetime
(from a pair of massive main sequence stars to compact binary formation and merger, or binary disruption, etc.)
sufficiently rapidly that large populations can be studied.

Such studies can be used to make significant progress on a wide range of topical questions, such as determining the
merger rate for compact objects and predicting the distribution of their observable
characteristics, and also to deepen our understanding of some of the physical processes
underpinning isolated binary evolution.

The specifics of some of the processes in the sequence of stellar interactions experienced by a binary are very poorly understood.
Such physical uncertainty can be captured by a parametrisation of the underlying astrophysics. These simulation
`hyperparameters' can then be varied in order to demonstrate the impact of different physical assumptions on our models. One of the principal goals of
COMPAS in the immediate future is to explore this hyperparameter space, both to understand the sensitivity of our simulations to changes
in different physical assumptions, and ultimately to infer constraints on their values from gravitational wave observations.

The systems of interest typically have
a low rate of occurrence relative to the total number of simulations. This means that evaluating a population of binaries large enough to form a significant sample of systems of interest
can take many hours of computation for any one set of hyperparameters. This makes the inverse problem of inferring hyperparameters from observations computationally intractable with a naive approach of densely covering the hyperparameter space with simulations. It is therefore necessary to minimise the computational cost associated with simulating binary populations.

In this paper we briefly describe some of the hyperparameters of interest in section~\ref{sec:hyperparameters},
and then present some preliminary work on tackling the computational intractability of the inverse problem in section~\ref{sec:emulation}.

\section{Simulation Hyperparameters}\label{sec:hyperparameters}

Although there are $\sim 10$ important hyperparameters governing isolated binary evolution,
as a first proof of concept we investigate two of the most important.

\subsection{Common Envelope Efficiency}

Common envelope events are arguably the least well understood phase of isolated binary evolution. These events are initiated
when one star expands beyond its Roche Lobe and begins dynamically unstable mass transfer. The material lost from the donor creates a so-called common envelope, encompassing both stars or stellar cores.  The stars spiral in inside the envelope, transferring orbital energy into the envelope. In some cases, enough energy
will be transferred into the envelope to expel it entirely; in others, energy loss will cause the binary to merge within the envelope.

The efficiency with which energy is transferred from the orbit to the envelope is highly uncertain, and is thus parametrised by a common envelope
efficiency parameter $\alpha_{\textrm{ce}}$, which is defined as the ratio of the binding energy of the envelope to the change in orbital energy of the binary (\cite[Webbink, 1984]{Webbink1984};
\cite[Ivanova, 2013]{Ivanova2013}):

\begin{equation}
  \alpha_{\textrm{ce}} = \frac{E_\textrm{env,Bind}}{E_\textrm{orb,fin} - E_\textrm{orb,ini}} \, .
\end{equation}

\subsection{Natal Kick Velocity Dispersion}

The asymmetric ejection of material in a supernova explosion can lead to a rocket effect, potentially
accelerating the star to velocities of order hundreds of km s$^{-1}$. This can have a significant effect on the eccentricity of the binary's orbit, which in turn
affects other evolutionary stages of the binary. Alternatively, large natal kicks can disrupt a binary
altogether, which affects merger rates and observed mass distributions.

The typical strength of such kicks is uncertain. Neutron star natal kicks are parametrised in terms of the dispersion term
$\sigma_\textrm{kick}$ in a Maxwellian distribution $P(v_\textrm{kick})$, from which any individual kick is randomly drawn:

\begin{equation}
  P(v_\textrm{kick}) = \sqrt{\frac{2}{\pi}} \; \frac{v_\textrm{kick}^2}{\sigma_\textrm{kick}^3} \exp\left(\frac{-v_\textrm{kick}^2}{2\sigma_\textrm{kick}^2}\right) \, ,
\end{equation}

with reduced kicks for black holes, depending on the amount of fallback
(\cite[Fryer \etal, 2012]{Fryer2012}).

\section{Binary Population Emulation}\label{sec:emulation}

Since the  parametrisations are strongly motivated by the underlying physics,
exploring and understanding them in depth will provide insight for
more detailed modelling of these processes.  However, as mentioned above, when approached naively, a full treatment of this problem is computationally intractable. It is simply not possible to explore the hyperparameter
space in detail by constructing a grid of populations, nor even by stochastically sampling
the hyperparameter space.

We are therefore using Gaussian process regression to yield robust and computationally efficient emulators to predict the outcome of population synthesis
simulations at unexplored points in the hyperparameter space, given a limited training set
of modelled populations at other points in hyperparameter space.

Gaussian process regression models the functions of interest as random variables that follow a multivariate Gaussian distribution. The number of input data points --- in our case, the number of points in hyperparameter space at which we generate simulations used to train the Gaussian process ---�� sets the number of dimensions in the Gaussian distribution.
The Gaussian process thus describes the functions of the hyperparameters $f(\mathbf{\theta})$ via a mean a mean function $\mu(\mathbf{\theta})$ and covariance function $\kappa(\mathbf{\theta,\theta'})$, where

\begin{equation}
  \mu(\mathbf{\theta}) = \left<f(\mathbf{\theta})\right>
\end{equation}
\begin{equation}
  \kappa(\mathbf{\theta,\theta'}) = \left< \left(f(\mathbf{\theta}) - \mu(\mathbf{\theta})\right) \left(f(\mathbf{\theta'}) - \mu(\mathbf{\theta'})\right)\right>
\end{equation}

(see \cite[Rasmussen \& Williams 2006]{Rasmussen_Williams_2006} for a more detailed introduction)

It is typical to take a centred distribution over functions, so that the mean function is zero. The covariance function is then a representation
of how the outputs behave with respect to differences in the inputs. Any positive definite function that is symmetric in $\theta$ and $\theta'$ is a valid
choice of covariance function. However, for simple problems by far the most widely used is the squared exponential covariance kernel:

\begin{equation}
  \kappa(\mathbf{\theta,\theta'}) = \sigma^2 \exp\left(-\frac{1}{2}\left|(\theta - \theta')^T \mathbf{g^{-1}} (\theta - \theta')\right|^2 \right) \, ,
\end{equation}

where $\mathbf{g}$ is a metric representing the length scale over the hyperparameters $\theta$, and $\sigma$ represents the variance of the output.  We treat $\sigma$ and the diagonal elements of $\mathbf{g}$ as free parameters.
It is these free parameters which are optimised with respect to observed input-output pairs.  We fixed them by maximising the  standard Gaussian likelihood as given in
(\cite[Rasmussen \& Williams 2006]{Rasmussen_Williams_2006}).
We optimised the free parameters $\sigma$ and $\mathbf{g}$
for each of our covariance functions via the L-BFGS gradient-based algorithm
(\cite[Nocedal \& Wright 1999]{Nocedal_Wright_1999}).
The resulting Gaussian process can then be used to make probabilistic predictions for function values at locations in hyperparameter space
where we have not performed any simulations.

We have focussed on two representative features of our populations. The first is the
fraction of simulated systems that form a system of interest, namely a compact object binary which will inspiral
and coalesce within a Hubble time (i.e., potential aLIGO sources). The second is the
distribution of chirp masses of systems of interest. The chirp mass is a particular
combination of the masses $M_{1,2}$ of the two compact objects which is well measured:

\begin{equation}
  M_{c} = \frac{\left(M_1 M_2\right)^{\frac{3}{5}}}{\left(M_1 + M_2\right)^{\frac{1}{5}}}\, .
\end{equation}

We ran an $11\times 11$ grid of populations at different points in the $2$-dimensional hyperparameter space of $\alpha_{\textrm{ce}}$ and $\sigma_\textrm{kick}$,
each consisting of $3$ million simulated binaries. We then held back a validation set of $15$ randomly chosen
populations, and used the remaining $106$ populations as a training set. Our goal was to predict both the merger rate
and the distribution of chirp masses in the validation set.

The chirp mass distributions are represented by a binned histogram, using the same bin locations and widths for every population. We then reduce the dimensionality of the histograms by employing a singular value decomposition to the matrix with one row per population and one column per chirp mass bin.
We used $80$ equally sized bins in chirp mass in the range $M_c = [0,40]M_{\odot}$, which we then projected on to a basis of the $10$ principal components. We found that the increase in predictive performance
when using more principal components was negligible.

We then trained separate Gaussian process emulators for each of these components, and one for the fraction of systems of interest. We used the \texttt{GPy} module\footnote{\textit{http://github.com/SheffieldML/GPy}}
in Python to perform all operations with our emulators.

\begin{figure}[h]
\begin{center}
 \includegraphics[width=4in]{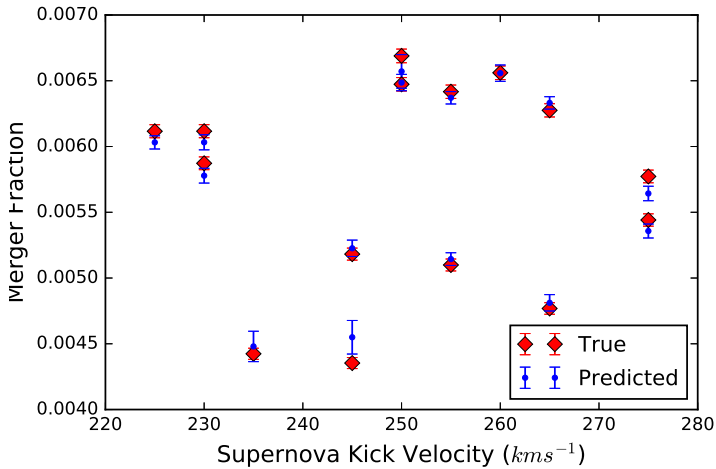}
 \caption{Comparison of the simulated and predicted fractions of merging binary black holes out of all simulated binaries at the validation points.  Where more than one point lies at the same
 value of the kick velocity, the validation hyperparameters have different values of common envelope efficiency. Error bars on the ``true'' values arise from counting statistics
 whereas errors reported on the predictions are the uncertainties reported by the Gaussian process regression.}
   \label{fig:ratesResults}
\end{center}
\end{figure}

\begin{figure}[h]
\begin{center}
 \includegraphics[width=5in]{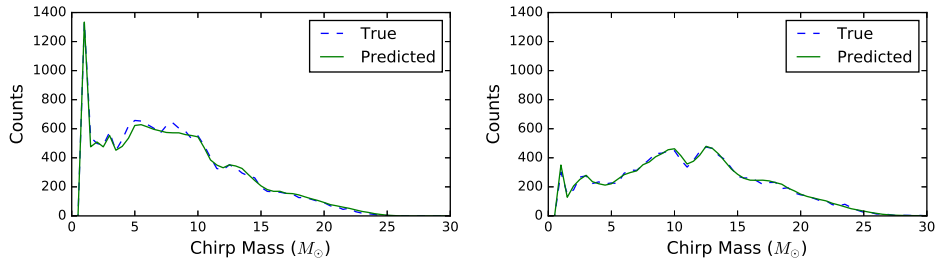}
 \caption{Results for two particular chirp mass distributions from the validation set, with hyperparameter values $(\alpha_{ce} = 1.5, v_{kick} = 250kms^{-1})$ and $(\alpha_{ce} = 0.7, v_{kick} = 255kms^{-1})$ respectively. These are representative of the emulator performance for all distributions in the validation set.}
   \label{fig:mChirpResults}
\end{center}
\end{figure}

We accurately predict both
the fraction of compact objects of interest (figure \ref{fig:ratesResults}) and the distribution of chirp masses (figure \ref{fig:mChirpResults}) using Gaussian process regression. Since, once trained, the Gaussian
process can predict these quantities very quickly, this work will allow us to tackle
the inverse problem of isolated binary population synthesis at a significantly reduced
computational cost.

There are several potential areas of further study arising from the work presented here.  Our work will need to be extended to higher dimensions, so that we can explore other binary evolution hyperparameters.  This may require modifications to the covariance function kernel, such as the use of a non-diagonal metric $\mathbf{g}$ or the joint modelling of all functions describing the population distribution.  Future improvements should also incorporate the statistical uncertainty in the Monte Carlo population models directly in the Gaussian process likelihood function.  A proper treatment of the input uncertainty would make it possible to compute a meaningful posterior distribution on the Gaussian process free parameters rather than just a maximum likelihood fit.  Finally, Gaussian process emulators could be used to determine where in the hyperparameter space future models should be evaluated in order to maximally improve the emulator quality at a minimal computational cost.


\begin{thebibliography}{}

\bibitem[Abadie \etal\ (2010)]{Abadie_etal2010}
{Abadie \etal} 2010, \textit{Classical and Quantum Gravity}, 27, 173001

\bibitem[Abbott \etal\ (2016a)]{Abbott_etal2016a}
{Abbott \etal} 2016a, \textit{Phys. Rev. Lett.}, 116, 061102

\bibitem[Abbott \etal\ (2016b)]{Abbott_etal2016b}
{Abbott \etal} 2016b, \textit{Phys. Rev. X}, 6, 041015

\bibitem[Abbott \etal\ (2016c)]{Abbott_etal2016c}
{Abbott \etal} 2016c, \textit{ApJ. Lett.}, 833, L1

\bibitem[Fryer \etal\ (2012)]{Fryer2012}
{Fryer \etal} 2012, \textit{ApJ}, 749, 91

\bibitem[Ivanova \etal\ (2013)]{Ivanova2013}
{Ivanova \etal} 2013, \textit{A\&AR}, 21, 59

\bibitem[Nocedal \& Wright (1999)]{Nocedal_Wright_1999}
{Nocedal \& Wright} 1999, \textit{Springer}

\bibitem[Rasmussen \& Williams (2006)]{Rasmussen_Williams_2006}
{Rasmussen \& Williams} 2006, \textit{MIT Press}

\bibitem[Stevenson \etal (2017)]{Stevenson_etal_prep}
{Stevenson \etal} 2017, \textit{Nature Communications}, 8, 14906

\bibitem[Webbink (1984)]{Webbink1984}
{Webbink} 1984, \textit{ApJ}, 277, 355

\end{thebibliography}
\end{document}